\documentclass[runningheads]{llncs}
\usepackage[T1]{fontenc}
\usepackage{graphicx}
\usepackage{tabularx}
\usepackage{todonotes}
\usepackage{url}
\usepackage[nolist]{acronym}
\usepackage{amsmath}
\usepackage{amssymb}
\newcommand{\R}{\mathbb{R}}
\newcommand{\set}[1]{\left\{#1\right\}}
\newcommand{\abs}[1]{\left|#1\right|}
\newcommand{\powerset}[1]{\mathcal{P}(#1)}

\begin{document}
	\title{Incentive-Based Software Security: Fair Micro-Payments for Writing Secure Code}
	\titlerunning{Fair Micro-Payments for Writing Secure Code}
	%
	\author{Stefan Rass\inst{1,2}\orcidID{0000-0003-2821-2489} \and
		Martin Pinzger\inst{3}\orcidID{0000-0002-5536-3859}}
	\authorrunning{S. Rass, M. Pinzger}
	%
	\institute{Johannes Kepler University, LIT Secure and Correct Systems Lab, Altenbergerstra{\ss}e 69, 4040 Linz, Austria
		\email{stefan.rass@jku.at}\\
		\url{http://www.jku.at/secsys} \and
		Institute for Artificial Intelligence and Cybersecurity, Alpen-Adria-University Klagenfurt, Universit\"atsstrasse 65-67, 9020 Klagenfurt, Austria
		\and
		Department of Informatics Systems, Alpen-Adria-University Klagenfurt, Universit\"atsstrasse 65-67, 9020 Klagenfurt, Austria}
	\maketitle              
	\begin{abstract}
		
		We describe a mechanism to create fair and explainable incentives for software developers to reward contributions to security of a product. We use cooperative game theory to model the actions of the developer team inside a risk management workflow, considering the team to actively work against known threats, and thereby receive micro-payments based on their performance. The use of the Shapley-value provides natural explanations here directly through (new) interpretations of the axiomatic grounding of the imputation. The resulting mechanism is straightforward to implement, and relies on standard tools from collaborative software development, such as are available for \texttt{git} repositories and mining thereof. The micropayment model itself is deterministic and does not rely on uncertain information outside the scope of the developer team or the enterprise, hence is void of assumptions about adversarial incentives, or user behavior,  up to their role in the risk management process that the mechanism is part of. We corroborate our model with a worked example based on real-life data.
	\end{abstract}
	\begin{acronym}
		\acro{CVE}{Common Vulnerabilities and Exposures}%
		\acro{CVSS}{Common Vulnerability Scoring System}%
		\acro{LDAP}{Lightweight Directory Access Protocol}%
		\acro{GDPR}{General Data Protection Regulation}%
		\acro{CC}{Common Criteria}%
	\end{acronym}

	\section{Introduction}
	
	Security has the unfortunate fate of not generating revenue by itself, but rather preventing damage at additional cost. As such, it does not necessarily ``add'' to the functionality of a system, but only protects it from malfunctions. Consequently, people may take considerably less satisfaction from implementing a security mechanism, since the system is working before and after, with no visible improvement other than increased robustness and security.
	
	The problem of reluctancy to design a product not only for functionality, but also for safety has a long history of research, including sophisticated game theoretic analysis using signalling games. Past such work found that companies may not necessarily have natural incentives to invest in security, for example, if customers are unwilling to pay the higher price for the more secure product \cite{RePEc:aea:aecrev:v:85:y:1995:i:5:p:1187-1206}.
	
	Investments in security may also be due to obligations of independent auditing, legal or (security) standard compliance. Frameworks like the \ac{CC} \cite{ccconsortiumCommonCriteriaInformation2018} or the ISO 27k family \cite{organisationISOIEC270002016} provide well formalized workflows along which security of a product, including its production process, can be established.
	
	These standards depend on developer teams to support the additional efforts imposed by a security-by-design paradigm, which we propose a (monetary) incentivization mechanism in this work. This is a form of incentive-based security \cite{huangRepChainReputationBasedSecure2021,liCreditCoinPrivacyPreservingBlockchainBased2018}, and relies on the hypothesis that people act rational towards maximizing their own utility. Research has found strong evidence against such a general utility maximization \cite{starmerDevelopmentsNonExpectedUtility2000,tverskyRationalChoiceFraming1989,tverskyFramingDecisionsPsychology1985}, which may be attributed not to a flaw in the general logic of utility maximization, but rather to a mere mistake in how the utilities are modeled. Indeed, research about bounded rationality has identified a considerable lot of reasons why people do not maximize a presumed utility expressed as a real-valued and continuous function, but rather consider multiple dimensions and uncertainty in their decisions. Hence, to create an easily perceptible incentive to maximize as one's own utility, we propose a pragmatic and straightforward mechanism, which is \emph{additional salary for people that implement security on top of the product's basic functionality}.
	
	Our mechanism employs cooperative game theory, specifically the Shapley-value to measure how much a member of a team contributes to the team goals, and from this, determines the additional payoffs to this person on top of the regular salary. For such a mechanism to work, we need mechanisms to keep track of people's contributions and to relate and assess them in light of security requirements to be (not) fulfilled by a person's actions. We will rely on a combination of code repository features and risk management processes for this purpose, to define a coalitional payoff function that we can use to reward team members in a fair manner. Our solution is cooked from the following ingredients:
	\begin{itemize}
		\item Risk management processes to systematically identify threats, countermeasures and implied conditions or test cases to verify the countermeasures as being (correctly) implemented in a software.
		\item Repository mechanisms that allow an attribution of certain changes to individual people, to verify which and whose contributions accomplished security requirements as found in the risk management.
		\item Sharing of the gains that the whole developer team earned among the team members in a systematic and ``fair'' way. To this end, the Shapley-value, specifically its axiomatic characterization will play a central role, since we can recognize several of its properties as naturally interpretable and useful for our purpose of incentivization.
	\end{itemize}
	We remark that the usual assumption of a ``grand coalition'' to form, which underlies the use of the Shapley-value in many instances, is automatically satisfied in our setting: we consider the entire developer team of an application as one (big) coalition of players working together against another party, which is the adversary. As such, we do not need to assume psychological or other mechanisms to yield the formation of a grand coalition, since this group is naturally found in the developer teams of software products.

\section{Related Work}\label{sec:related-work}
The challenge of incentivizing developers of software to care more for security has long been recognized. The early work of \cite{augustDesigningUserIncentives2014} states that ``the design of an incentives-based approach to improve cybersecurity is a difficult task because the level of risk that realizes on a given system or network is a complex outcome of the behaviors of many stakeholders: government, critical infrastructure providers, technology producers, malicious (`black hat') hackers, and users''. This motivates a change to another, more intrinsic, mechanism that quantifies security contributions exclusively on the actions of the developers, which we can measure (unlike all the other ``variables'' mentioned above).

The idea of incentivizing developers to declare security as its own goal is contrasted by the dual approach of transferring risk. In special contexts such as IoT, this can mean traditional mechanisms of risk transfer to third parties (such as insurances) \cite{adatEconomicIncentiveBased2018}, but early game-theoretic treatments of company liabilities for insufficient security (modeled by signalling games) have shown that investments into security can correlate with pricing in a way that can create even the opposite incentive of \emph{not} investing in security \cite{RePEc:aea:aecrev:v:85:y:1995:i:5:p:1187-1206}. The work of \cite{haldermanStrengthenSecurityChange2010} provides eloquent discussions about the needs for the right incentives, and argues for incentivization using transparency and liability mechanisms. In light of aforementioned research, especially liabilities may game-theoretically induce unwanted effects, so that the problem seems to require other mechanisms (one of which we propose in this work). Possible negative effects of market-based incentives for security were independently also found by \cite{eetenEconomicsMalwareSecurity2008}.

Incentivization mechanisms for security do not need to rely on the developers themselves, but can equally well root in the user group. Bug bounty programs are a common instance here, offering rewards for the discovery of threats. Our work addresses the aftermath thereof, creating incentives for those who contribute to mitigating risks. The work of \cite{augustNetworkSoftwareSecurity2006} provides such a treatment, discussing different patching strategies, and \cite{garayRationalProtocolDesign2013,gordonRationalSecretSharing2006,kawachiGeneralConstructionsRational2017,huangRepChainReputationBasedSecure2021,liCreditCoinPrivacyPreservingBlockchainBased2018} use incentivization mechanisms to encourage honest behavior in cryptographic protocols and blockchains. Similarly, \cite{liuIncentivebasedModelingInference2005a} use non-cooperative game theory to study incentives for attackers. 

Applications of the Shapley-value have been described for software alliances \cite{lvResearchProfitDistribution2013} and to share threat intelligence \cite{xieImprovedShapleyValue2020}, which suggests using similar concepts to incentivize developer teams as done in this work. 

\section{Preliminaries}

\subsection{Collaborative Software Development}
We assume that a software development team consists of a finite set $N$ of persons. Furthermore, let there be a collaboratively used code repository (e.g., \texttt{git}), to which all developers can commit and push changes, possibly with mandatory pull requests. The versioning mechanism and reviews of pull requests before merging changes into a new version, are the mechanism that we can use to ``quantify'' people's contributions to security, and based on this, provide a payback mechanism to reward people for contributions that improved the security.

\subsection{Qualitative Risk Management}\label{sec:risk-management}
Let there be a risk management process, such as ISO 27k or similar, running in the background, along which a set of threats was identified and quantified in terms of likelihoods and impacts. Experience from practice has shown that it is often not advisable to hope for much accuracy in either, the likelihoods or damages, and categorical specifications of how likely an event may be, or giving damage ranges instead of exact estimates is widely adopted practical method.

Speaking about damages, one may think of a certain threat $T_i$ to cause some practically unknown (and hard to accurately anticipate) damage $d_i$, which, nonetheless, lies in some range that we are rather sure about. That is, we may -- with reasonable certainty -- partition the entire range of possible damages into a finite number\footnote{generally, such partitions would be non-equidistant, since the perception of ``low'' and ``high'' strongly depends on the context and even the person. This applies to both, likelihood and impact scales; see \cite{starmerDevelopmentsNonExpectedUtility2000} for a deeper discussion} of nonempty disjoint intervals $D_i=[d_{i*},d_{i+1}^*)$ starting from $d_{1*}=0$ (no damage) and for $i=1,2,\ldots$ until $d_{i+1}=\infty$ after a finite number of steps. This delivers an ordered set of damage categories $\mathcal{D}=\set{D_1<D_2<\ldots <D_{i_{\text{max}}}=[d_{i_{\text{max}}*},\infty)}$. If we classify a threat to cause ``damage of category $D_i$'', we then mean that the actual damage will lie within the bounds that $d_{i*}$ at least and $d_i^*$ at most, with both numbers being known. This is practically convenient, since security risks do not necessarily have a ``ground truth'' (like coming from a physically measurable process). Even if exact statistical models exist, finding them is not necessarily feasible in practice. The use of qualitative risk scales, here represented by intervals, has a twofold benefit: it (i) saves us from complicated and hard to justify distributional assumptions about the randomness of the damage, and (ii) avoids estimates for which no ultimate precision is possible, since it is nonetheless a \emph{conjectured} damage that \emph{may}, but not need to, happen in future.

Similarly, we can quantify likelihoods by partitioning the unit interval $[0,1]=\bigcup_{j=1}^c [\ell_{j*},\ell_{j+1}^*)$ into a finite number of $c$ categories, with $\ell_{1*}=0$ and $\ell_c^*=1$. The scale is thus again an ordered set of intervals $\mathcal{L}=\set{L_j=[\ell_{j*},\ell_{j+1}^*)}_{j=1}^c$, where a concrete likelihood $\Pr(T)$ for some threat $T$ being specified qualitatively as $\Pr(T)=L_i$, meaning that the actual probability is somwhere between $\ell_{i*}$ and $\ell_i^*$.

Like with the damage, this saves from unreliable estimates for likelihoods of events that were so rare or so little reported that no robust statistics about relative incident frequencies can be compiled. Nonetheless, we must not ignore security risks because we do not have information or data, and will need to resort to subjective assessments in these cases. Qualitative scales are a convenient method of assessment here, well justified in  \cite{munchWegeZurRisikobewertung2012}.

Practically, both scales, for impact and likelihood, are kept small, with typically 3 to 5 categories, named as ``low'', ``medium'', ``high'' or similar. For subjective assessments, it is sometimes advised to work with an even number of categories to do not let people choose the ``middle'' category (if the scale has 3 or 5 categories) in case that they do not know, so that they must specify at least a tendency.

The interpretation of categories as intervals is particularly important as an assumption for our upcoming incentivization mechanism in Section \ref{sec:incentivization-mechanism}.

\subsection{Coalitional Gains and Individual Rewards}
For the developer team, represented as the set $N$, let us write $v(N)$ for the total gain in security that they can cooperatively accomplish. A core part of this work is to describe how to define and compute $v(N)$, but let us for the moment assume that we have such a function. Taking its value directy as the team gain, the Shapley-value provides a method of returning these profits to group members.
\begin{definition}[Shapley-Value]
Given a function $v\colon\powerset{S}\to\R$ be a function with $v(\emptyset)=0$, and with $S$ as a set of players. Let a coalition $C\subseteq N$ of players be assigned the total gains $v(C)$. The payback to an individual player $i\in C$ is the \emph{Shapley-value}, given by
\begin{equation}\label{eqn:shapley}
	\phi_i(v) = \sum_{S\subseteq N\setminus\set{i}}\frac{\abs{S}!(\abs{N}-\abs{S}-1)!}{\abs{N}!}(v(S\cup\set{i})-v(S))
\end{equation}
\end{definition}
We propose using the Shapley-value for a fair revenue of efforts towards increased security of a software product, based on the hypothesis that the coalitional payoff $v(S)$ measures how much the team $S$ accomplished towards increasing security. The inner term $v(S\cup\set{i})-v(S)$ then corresponds to how much additional revenue a new team member $i$ would bring to the existing team $S$. Some code repositories like \texttt{git} have useful features to aid the computation of $v(S\cup\set{i})-v(S)$ rather directly, which we will describe in Section \ref{sec:git-cherrypicking}, Section \ref{sec:git-commits} and a worked example in Section \ref{sec:worked-example}. The Shapley-value has an even more appealing axiomatic characterization. Let $v(N)$ be what team $N$ gains as bonus for efforts towards making a product more secure. The incentivization mechanism is to pay back the gains $v(N)$ to the individual team members, each of which gets a share $\phi_i(v)$, and the Shapley-value is provable to be the only such assignment with the following properties:
\begin{enumerate}
	\item \emph{Efficiency}: the entire value $v(N)$ will be distributed among the team members, i.e.,
	\[
		\sum_{i\in N}\phi_i(v)=v(N).
	\]
	That is, if $N$ is the full developer team, there will be no ``leftovers'' that a team megmber may claim in addition.
	\item \emph{Symmetry}: if two people contribute the same amount to the team, then they get equal payments. Formally, if $v(S\cup\set{i})=v(S\cup\set{j})$ for all subsets $S\subseteq N\setminus\set{i,j}$ that contain neither $i$ nor $j$, then $\phi_i(v)=\phi_j(v)$.
	\item \emph{Null player}: if a person contributes nothing to a team, irrespectively how the team is set up, then this person receives nothing. Formally, if $v(S\cup\set{i})=v(S)$ for all sub-teams $S\subset N$ that do not contain $i$, then $\phi_i(v)=0$.
	
	This does allow the possibility that a person may receive some bonus if it joins a particular team $S_0$, for which then $v(S_0\cup\set{i})>v(S_0)$, in which case $\phi_i(v)>0$ will be paid to this person. However, only in case that a person is ``universally useless'', there will be no payment. From an incentivization perspective, this creates a benefit for a person to either bring in something to the current team, or find itself another team to which one can be an asset.
	
	\item \emph{Linearity}: for two functions $v,w$, we have $\phi_i(v+w)=\phi_i(v)+\phi_i(w)$.
	
	For software security team work, this translates into the simple fact that if a person's contributed part to a product or a software is used and useful in multiple projects, quantified by distinct functions $v,w$, then the bonuses gained from these multiple contributions add up. Despite software engineering best practices typically advising people should not work on more than one project at a time, this improves the incentive for people if their work has wider applicability. 
	
	The case of a person contributing the same piece of source code to several teams identically would, by linearity, mean that this person receives the respective payment multiple times. Interpreting such a contribution as ``more important'' because it helps several developer teams at the same time, proportionally increasing the payment for it seems admissible.
\end{enumerate} 
Computing the Shapley-value is, today, made easy by a variety of tools supporting this, such as \cite{cano-berlangaEnjoyingCooperativeGames2015,saavedra-nievesGameTheoryAllocationToolsCalculating2016}.

\subsection{Team Contributions via Risk Reductions}
To define a function $v$ that measures how much the team creates as revenue by collaboration towards security, we continue the risk management process from Section \ref{sec:risk-management}.

Once a list of threats like abstractly shown in Table \ref{tbl:threat-enumeration} has been identified and quantified according to pre-defined scales as described in Section \ref{sec:risk-management}. We hereafter assume that the list of threats is to be covered entirely, i.e., \emph{all} respective controls have to be implemented. The creation of this list is part of an (outer) risk management process, inside which our incentivation mechanism is embedded. Along this process, excluded from the scope of this work, matters of mutual dependence between controls (ranging from independence up to mutual exclusion or substitution) are all covered during the compilation of the threat and countermeasures list.



\begin{table}
	\centering
	\caption{Threat enumeration for risk management}\label{tbl:threat-enumeration}
	\begin{tabular}{|c|c|c|c|}
		\hline 
		Threat & Likelihood $\in\mathcal{L}$ & Impact $\in\mathcal{D}$ & Risk = Likelihood $\times$ Impact\tabularnewline
		\hline 
		\hline 
		$T_{1}$ & $\ell_{T_{1}}$ & $d_{T_{1}}$ & $r_{T_{1}}=\ell_{T_{1}}\times d_{T_{1}}$\tabularnewline
		\hline 
		$T_{2}$ & $\ell_{T_{2}}$ & $d_{T_{2}}$ & $r_{T_{2}}=\ell_{T_{2}}\times d_{T_{2}}$\tabularnewline
		\hline 
		$\vdots$ &  &  & \tabularnewline
		\hline 
		$T_{n}$ & $\ell_{T_{n}}$ & $d_{T_{n}}$ & $r_{T_{n}}=\ell_{T_{n}}\times d_{T_{n}}$\tabularnewline
		\hline 
	\end{tabular}
\end{table}

The overall risk is simply the sum of all individual risks, and reducing this sum to zero is the most that the developer team can accomplish, although this theoretical optimum is in most pratical cases unachievable. Still, the process yields a quantity from which we can define the team contributions later, 
\begin{equation}\label{eqn:risk-formula}
\sum_{i=1}^n \ell_{T_i}\cdot d_{T_i} = \sum_{i=1}^n \Pr(T_i)\cdot d_{T_i},
\end{equation}
in which $\Pr(T_i)$ is assumed to take ordinal values, according to risk management best practices \cite{munchWegeZurRisikobewertung2012}.

\subsection{Systematic Risk Reductions as Team Goals}

The actual team contribution is then defined as the amount by which the risk reduces thanks to the team's actions. For a verifiable such measurement, we need controls that the developer team can implement verifiably. For example, suitable controls are listed in companion catalogues to (many) risk management standards, and the effect of implementing a control can be twofold:
\begin{itemize}
	\item Reduction of damage $d_{T_i}$ in case that the threat $T_i$ becomes real. This may lead to some reduced damage $d_{T_i}'<d_{T_i}$. If the modeling admits the specification of some damage $X\sim F$ with probability distribution $F$ (e.g., an extreme value distribution or similar), one may take $d_{T_i}=E_F(X)$, respectively the conditional expectation $d_{T_i}'=E_F(X~|~\text{countermeasure})$, in a slight abuse of notation.
	
	\begin{example}[Backups]
		For backups, the full value of the data $d_{T_i=\text{data-loss}}$ is reduced to only much smaller residue loss $d'$ after having recovered a perhaps only slightly older version of the data.
	\end{example}

	\item Reduction of the likelihood $\ell_{T_i}$: Without countermeasures, we may model $\ell_{T_i}$ as unconditional probability $\ell_{T_i}=\Pr(T_i)$, or, in light of countermeasures, as conditional probability $\Pr(T_i~|~\text{countermeasure})$ with the additional assumption that $\Pr(T_i~|~\text{countermeasure})<\Pr(T_i)$ (to be proven in practice, but reasonably assumable since otherwise the countermeasure would be useless).
	
	\begin{example}[Encryption]
		Encrypting data does not avoid the event of eavesdropping, but substantially reduces the chances of information to leak out. Practically, under contemporary cryptographic security being correctly implemented, we may thus assume $\Pr(T_i=$~plaintext information leakage$)=p>0$, but $\Pr(T_i=$~plaintext information leakage$~|~$encryption in place$)\approx 0\ll p$.
	\end{example}
\end{itemize}
We remark that both, the (conditional) residual damage and conditional (residual) likelihood are to be understood as measures on an ordinal scale, and not with the usual interpretation as continuous quantities. While the latter interpretation is in no way incorrect from a theoretical perspective, risk management best practices advise to not quantify likelihood or damages on a continuous scale, but rather on ordinal scales \cite{munchWegeZurRisikobewertung2012}.

Returning to our risk formula, and taking a countermeasure $C_{i,j}$ (possibly among others) to be correctly implemented, the values in \eqref{eqn:risk-formula} become accordingly reduced, here denoted by the $'$ annotations and conditional probabilities, and called \emph{risk after mitigation} (in the risk management literature) 

\begin{align}
	\text{risk after}&\text{~countermeasure} = \sum_{i=1}^n \Pr(T_i~|~C_{i,j})\cdot\ell_i'\label{eqn:risk-after-mitigation}\\&<\sum_{i=1}^n \Pr(T_i)\cdot\ell_i = \text{risk without precautions}\label{eqn:risk-before-mitigation}
\end{align}

\begin{figure}
	\centering
	\includegraphics[width=\textwidth]{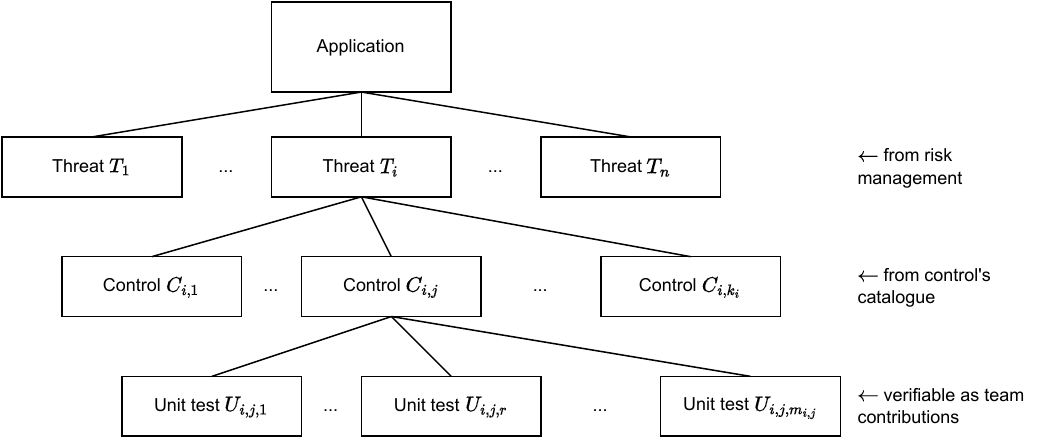}	
	
	\caption{Threat Tree: Threats, implying countermeasures, verifiable by certain required conditions}\label{fig:threat-tree}
\end{figure}

\begin{remark}
We herein do not consider new threats or vulnerabilities possibly induced by changes that a team applies to the code. In that case, and on a regular basis anyway, a reconsideration of the threat list (Table \ref{fig:threat-tree}) is required. Here, we assume that the changes of the developers will make the threat list monotonously become shorter, until a re-assessment is made after which we re-start with an updated new table. This table then contains possible new threats induced by the changes made in previous cycles, that most risk management standards prescribe anyway (such as ISO27k, for example).
\end{remark}



Now, among the finite total of $n$ threats, let the $i$-th threat have $k_i$ countermeasures associated with it. These may be systematically identified by various techniques, such as \ac{CC} \cite{ccconsortiumCommonCriteriaInformation2018}, secure coding guidelines \cite{howardWritingSecureCode2003} or catalogues of the ISO 27k standards \cite{isoISOIEC270022023}.

Likewise, let the $j$-th countermeasure $C_{i,j}$ be checked against a set of test cases, which can (but not need to be) unit tests or manual reviews (e.g., as described in standards like \ac{CC} in the auditing guidelines \cite{ccconsortiumCommonCriteriaInformation2012a}), here denoted by $U_{i,j,r}$ for the $r$-th among a total of $m_{i,j}$ checks conducted to verify that risk control $C_{i,j}$ was correctly implemented. The exact number of controls or verifiable conditions thereto is of no further interest in this work, except for the assumption that the numbers are finite and feasibly low to admit an automated checking.

Our incentivization mechanism will be based on how many (leaf) conditions in the threat tree from Figure \ref{fig:threat-tree} have been addressed by the team member's commits. This number is upper bounded by the number of leaves in the threat tree. The total contribution of a set $S$ of people from the developer team is thereby quantifiable as 
\begin{equation}\label{eqn:team-contribution}
	v(S)=\frac{\text{number of requirements addressed by members of }S}{\text{number of leaves in the threat tree}}\leq 1.
\end{equation}

\subsection{Computing $(v(S\cup\set{i})-v(S))$ by \texttt{git} Cherry Picking}\label{sec:git-cherrypicking}
To reward a team member $i$ for individual contributions, we need to compute the difference $v(S\cup\set{i})-v(S)$, which is almost directly supported by repository mechanisms. Namely, we can cherry-pick commits of a certain set of authors and carry over all these into a new branch, re-compile and verify the existing test-cases. More specifically, a \texttt{git} repository would let us compute the sought difference as follows: since we need to assume $i\in S$, as person $i$ has, as all others, contributed to the current status of the application, we put $S':=S\setminus\set{i}$ so that $S=S'\cup\set{i}$ and equivalently compute the difference $v(S)-v(S\setminus\set{i})$:
\begin{enumerate}
	\item Take the current status of the repository and compute $v_1 = v(S)$ using \eqref{eqn:team-contribution}. 
	\item Fix a point in time, represented by a certain commit (has) that we here call \verb|<baseCommit>| after which the new team contributions shall be quantified for micropayments (e.g., along an annual team appraisal or in other cycles).
	\item Create a branch starting from the starting point \verb|<baseCommit>|
	\item Cherry-pick all commits from the authors in the $S\setminus\set{i}$ into the new branch, rebuild the application and re-evaluate the value $v_2 = v(S\setminus\set{i})$ accordingly.
\end{enumerate}
Compute the expression $v(S'\cup\set{i})-v(S')=v_1 - v_2$ in  \eqref{eqn:shapley} to quantify each team member's contribution. 

Cherry picking may have difficulties in producing potentially dysfunctional or even non-compiling code, if the commits are such that leaving some out will leave other parts of a program not working. Although this is not necessarily a problem for the value computation as such (see Example \ref{exa:shapley-example} for a case where a team of two people have both made contributions which by themselves would be not working, but jointly do contribute to security), reviewed pull requests or manual checks of adherence to coding best practices may be another option.

\subsection{Computing $(v(S\cup\set{i})-v(S))$ by \texttt{git} Pull Requests}\label{sec:git-commits}

Pull requests are usually undergoing a check, up to a code review, by some authority that judges whether or not a feature has been correctly addressed and hence will be merged into the main branch. This authority would need to do decide if the pull request would still be accepted without certain commits by certain persons. In a simplified setting, we can collect all names from people who made commitments that were part of the pull request, let us call this set $P\subseteq N$. Then, we may put $v(P)=v(P'')=v(N)=1$ for all $P''\supseteq P$ and $v(P')=0$ for all $P'\subsetneq P$. Collecting this data from \texttt{git} repositories, such as \texttt{GitHub}, is further facilitated by existing data sources, such as \cite{gousiosGhtorrentGhtorrentOrg2020}, or repository mining tools, such as Perceval \cite{duenasPercevalSoftwareProject2018}.

\begin{example}\label{exa:shapley-example}
	Assume that, for example, two team members have jointly worked on a security feature, which fails if either of the two's contributions are omitted, but the payments may still be distributed fair among the players. To showcase this, let the team be only of three members $N=\set{1,2,3}$, and assume that only person 1 and person 2 together contributed something to security, while neither's commits alone would be working. The value assignment is thus:
	
	\begin{center}
		\begin{tabularx}{\textwidth}{|c|c|X|}
			\hline 
			Coalition & Value $v$ & Comment\tabularnewline
			\hline 
			\hline 
			$\emptyset$ & 0 & security does not come from nothing (no team member involved)\tabularnewline
			\hline 
			\{1\} & 0 & person 1 has made only partial contributions that did not by themselves
			add anything to security\tabularnewline
			\hline 
			\{2\} & 0 & person 2 has made complementary contributions to person 1, but also
			these were not enough for a working security feature\tabularnewline
			\hline 
			\{1,2\} & 1 & jointly, 1 and 2 have accomplished a security feature\tabularnewline
			\hline 
			\{3\} & 1 & person 3 has by oneself made a contribution to security\tabularnewline
			\hline 
			\{3,1\} & 1 & person 1 did not add anything to what person 3 did\tabularnewline
			\hline 
			\{3,2\} & 1 & person 2 was working together with person 1, but added nothing to
			what person 3 has done\tabularnewline
			\hline 
			\{3,1,2\} & 2 & the entire team addressed two security features\tabularnewline
			\hline 
		\end{tabularx}
	\end{center}

	The Shapley-values of the so-assigned coalitions are $\phi_1(v)=0.5=\phi_2(v)$ and $\phi_3(v)=1$. This is in alignment with the above findings, since person 3 has -- alone -- accomplished a certain security feature, while neither person 1 or person 2 can claim the credit for the second feature only for her/himself. However, their joint commitments satisfied another requirement that person 3 did not address, so their payment is 0.5 each, and half of what person 3 receives, who addressed the other feature alone. 
\end{example}

\section{Rewarding the Team Members}\label{sec:incentivization-mechanism}

The ``gain'' from having implemented all countermeasures is just the reduction of risk, namely value \eqref{eqn:risk-before-mitigation} minus value \eqref{eqn:risk-after-mitigation}, i.e.,
\begin{equation}\label{eqn:risk-delta}
	\Delta = \sum_{i=1}^n [\Pr(T_i)\cdot\ell_i-\Pr(T_i~|\text{~after mitigation})\cdot\ell_i'],
\end{equation}
where the sum runs over the list of threats, and $\ell_i'$ is the residual damage despite the countermeasure being correctly implemented.

The value $\Delta$ is, according to best practices mentioned before, usually a categorical value, representable as integer (ranks), but by itself meaningless as it is a mere score, and more precisely, a difference of scores.

A conversion of $\Delta$ back into monetary savings is not generally possible, but may in practice be based on the meaning of damage categories. However, since we initially defined the categories to be intervals, we can directly apply interval arithmetic to convert $\Delta$ into an interval, based on the ranges in which $\Pr(T_i), \Pr(T_i~|\text{~after mitigation}), \ell_i$ and $\ell_i'$ lie.

Evaluating \eqref{eqn:risk-delta} in terms of interval arithmetic (very well tool supported, e.g., \cite{sets1,sets2}), we get a range for $\Delta=[\delta_*,\delta^*]$ that now covers the \emph{expected improvement} by the actions of the team members. That is, recalling that \eqref{eqn:risk-after-mitigation} and \eqref{eqn:risk-before-mitigation} are just expressions of probabilities multiplied by outcomes, both are expected damages, and \eqref{eqn:risk-delta} is a difference of expectations. Since expectation is a linear operator, $\Delta$ is itself an expectation of the improvement in terms of risk.

This improvement is now easily convertible into a revenue for the team members, since their contribution saved the company from a loss of at least $\delta_*$ and at most $\delta^*$, according to their own risk management processes and categories.

At this point, it is the company's decision of how much of the savings, relative to the range $[\delta_*,\delta^*]$ they are willing to pay back to the team members as incentive to put effort on security. Let this amount be $B\in[\delta_*,\delta^*]$, then, based on the Shapley-value, each person of the team $N$ receives a share of 
\begin{equation}\label{eqn:final-payment}
	\psi(i)\propto \phi_i(v)\cdot B
\end{equation}
as incentivization (or bonus) for person $i$, proportional to the cost savings that the company thanks this person $i$ for. The payments to all persons can (for reasons to be discussed in Section \ref{sec:conclusion}) be made publicly and known to all team members. The size of the budget $B$ is here assumed to be ``reasonably small'', so that the company may not (want to) afford a designated person or team working exclusively on security. Hence, the mechanism is mainly intended for cases (or companies) who cannot afford extra staff designated to securing code, but has some budget to reward the developers themselves to take care of writing secure code.

\section{A Worked Example}\label{sec:worked-example}
To corroborate the model, we consider an example of a real-life project with actual (now resolved) vulnerabilities, indicated by \ac{CVE} numbers, rated by the \ac{CVSS}, and put in the context of a hypothetical company to use the model in their product development. This example is only intended to demonstrate how the model can be instantiated, especially to show where to get the parameters from, and how to work with data and information available in a practical environment.

Our software example is Apache Streampark \cite{apachefoundationApacheStreamParkIncubating2023}, whose version 1.0.0 had the followling reported vulnerabilities \cite{informationtechnologylaboratoryNVDHome2023}, with shortened descriptions given in Table \ref{tbl:vulnerabilities}. We take these vulnerabilities as system conditions that enable the \emph{same} (single) threat $T_1 := $~``loss of customer data records''. As in Figure \ref{fig:threat-tree}, the three possibilities to mount an attack induced by threat $T_1$ entail three respective controls $C_{1,1},C_{1,2}$ and $C_{1,3}$, directly based on fixing the vulnerabilities accordingly. 

\begin{table}
	\centering
	\caption{Reported Vulnerabilities with Ratings}\label{tbl:vulnerabilities}
	\begin{tabular}{|p{3cm}|p{6cm}|c|}
	\hline 
	Vulnerability \cite{informationtechnologylaboratoryNVDHome} & Shortened description & \ac{CVSS}v3.1 Score\tabularnewline
	\hline 
	\hline 
	CVE-2022-46365 ($\to$ control $C_{1,1}:=$ ``verify user names'') & unverified user-name (this will allow malicious attackers to send
	any username to modify and reset the account) & 9.1 (critical)\tabularnewline
	\hline 
	CVE-2022-45802 ($\to$ control $C_{1,2}:=$ ``verify file types'') & uploaded file types unverified (upload some high-risk files, and may
	upload them to any directory) & 9.8 (critical)\tabularnewline
	\hline 
	CVE-2022-45801 ($\to$ control $C_{1,3}:=$ ``LDAP query sanatization'') & LDAP injection vulnerability (upload some high-risk files, and may
	upload them to any directory) & 5.4 (medium)\tabularnewline
	\hline 
\end{tabular}
\end{table}
We simplify the risk management process in this example, by equating the \ac{CVSS} risk scoring directly to the business impact levels (although the two are defined differently). The next table gives the definition of \ac{CVSS} ratings and business impact levels, both defined on the same scale. The ranges of loss are based on the 2015 Verizon Data Breach Investigation Report \cite[Fig. 23]{widup2015VerizonData2015}: the data given below is taken from this document, which in turn is based on a statistical analysis of cyber-insurance data related to losses due to leakage of customer records. Specifically, the ranges given in our example refer to the lower- and upper bounds of the union of 95\% confidence intervals around the average losses that enterprises suffered due to the loss of customer data (in ranges of up to 100 records (category ``low''), 1000\ldots 10\,000 records (category ``medium''), 100\,000\ldots 1\,000\,000 records (category ``high'') and 10 million up to 100 million records (category ``critical'')). In our example, we will assume that any of the three vulnerabilities of the software could lead to loss of customer data, whose losses can be quantified based on the data from the cited report\footnote{We emphasize that, consistently with what the authors of \cite{widup2015VerizonData2015} say, that these numbers are merely indicative and cannot accurately reflect any actual losses in reality; nonetheless, they shall serve as an illustration of how losses may be quantified. Any such quantification is, however, to be taken with at least a grain of salt. The 2022 version of the same report no longer contains likewise figures.}.

\begin{table}
	\centering
	\caption{Impact definitions (for our example company)}\label{tbl:business-impacts}
	\begin{tabularx}{\textwidth}{|c|c|X|X|}
		\hline 
		Level & \ac{CVSS}v3.1 condition \cite{firstforumofincidentresponseandsecurityteamsCVSSV3Specification} & business impact condition & estimated losses (based on \cite{widup2015VerizonData2015})\tabularnewline
		\hline 
		\hline 
		none & score 0.0 & no loss of customer data & none \tabularnewline
		\hline 
		low & $0.1\leq$ score $\leq3.9$ & up to 100 customer records lost & between \$18,120 and \$35,730\tabularnewline
		\hline 
		medium & $4.0\leq$ score $\leq6.9$ & up to 10 000 customer records lost & between \$52 260 and \$223 400\tabularnewline
		\hline 
		high & $7.0\leq$ score $\leq8.9$ & up to 1 mil. customer records lost & between \$ 366 500 and \$ 1 775 350\tabularnewline
		\hline 
		critical & $9.0\leq$ score $\leq10.0$ & up to 10 mil. customer records lost & between \$2 125 900 and \$15 622 700\tabularnewline
		\hline 
	\end{tabularx}
\end{table}

Based on this data, let us now instantiate the variables to appear in the computation of $\Delta$ in \eqref{eqn:risk-delta}:

For the \emph{impact assessment}, we simply equate the \ac{CVSS} ranking with the business impact, assigning the vulnerabilities the same impact categories as noted in the description table above, giving the loss range $\ell_1 = [\$~2\,125\,900, \$~15\,622\,700]$.

For the \emph{likelihood assessment}, since the vulnerabilities have been reported over a publicly accessible database, we assign all vulnerabilities the exploit likelihood ``high'', assuming that a known exploit will be used eventually, putting $\Pr(T_1)=1$.

The \emph{risk} for the three vulnerabilities, by the usual ``risk $=$ impact $\times$ likelihood'' formula symbolically becomes $\max\{$``medium'', ``critical''$\}\times $``high''$=$ ``critical'', applying the maximum principle of system security (letting the highest impact determine the overall risk of the system).

As of version 2.0.0 of Apache Streampark, all three vulnerabilities are no longer relevant, which we interpret as being verifiably fixed (hence implicitly indicating that there have been respective test-cases, e.g., codes to demonstrate the vulnerability in version 1.0.0, and showing that the same code\footnote{The \ac{CVSS} scoring could indicate such a demo code to exist, although no such indication is given for the vulnerabilities reported here; still the attack complexity was rated as ``low'' in all three cases} no longer works in version 2.0.0, which would be unit tests $U_{1,1,1}, U_{1,2,1}$ and $U_{1,3,1}$ associated with the controls $C_{1,1},\ldots,C_{1,3}$; cf. Figure \ref{fig:threat-tree}). Hence, the probability of an exploit after mitigation is $\Pr(T_1~|~$after mitigation$)=0$.

Evaluating \eqref{eqn:risk-delta} leads to the following losses being avoided since the vulnerabilities have been fixed. Using interval arithmetic \cite{sets1}, we compute $\Delta = \ell_1 = [\$~2\,125\,900,\$~15\,622\,700]$.

To answer how much of this range our example company would be willing to pay back to the people having fixed the problems, we take a look at the fine that our enterprise would face in case of customer data breaches. According to the \ac{GDPR} \cite[Art.83(5)]{european_commission_regulation_2016}, up to 20 million Euros or, in the case of an undertaking, up to 4\% of the worldwide annual turnover of the preceding financial year (whichever is higher), would be possible to pay. We assume that a (small) fraction of the minimum losses (not even fines) would be allocated as budget to incentivize developers. So, let us take a fraction of $1\%$ of the minimum loss $\min\ell_i=\$~2\,125\,900$ to define the budget $B := 0.01\cdot 2\,125\,900=\$~21\,259$ for the developer incentivization (the 1\% fraction is the proportionality factor appearing in \eqref{eqn:final-payment}).

To share this budget according to our Shapley-value based model, assume that the three vulnerabilities were collaboratively fixed by three developers, named Alice, Bob and Carol, who worked together as follows, documented by (here entirely hypothetical) respective \texttt{git pull} requests\footnote{Any coincidental match to the real people responsible for the fixing of the vulnerabilites in reality are unintended; we stress that the real credit goes to real people and their hard work on the project; our assumed pattern of who fixed what is entirely artificial and for illustration only}:

\begin{itemize}
	\item Alice: rectified CVE-2022-46365, and CVE-2022-45801
	\item Bob: rectified CVE-2022-45802
	\item Carol: rectified CVE-2022-45802, CVE-2022-45801
\end{itemize}

With this knowledge, it is straighforward to define the coalitional payoffs as the number of vulnerabilities fixed per subset of $\set{A,B,C}=\{$Alice, Bob, Carol$\}$, as listed in Table \ref{tbl:example-contributions}.

\begin{table}
	\centering
	\caption{Defining the coalitional payoffs based on the number of fixed vulnerabilities}\label{tbl:example-contributions}
	\begin{tabular}{|c|p{6cm}|p{2.5cm}|}
		\hline 
		Team & addressed vulnerabilities & accomplishment $v$\quad (= count)\tabularnewline
		\hline 
		\hline 
		$\emptyset$ & none & 0\tabularnewline
		\hline 
		$\{A\}$ & CVE-2022-46365 & 1\tabularnewline
		\hline 
		$\{B\}$ & none (fixes completed only in collaboration with Alice or Carol) & 0\tabularnewline
		\hline 
		$\{C\}$ & none (fixes completed only in collaboration with Alice or Bob) & 0\tabularnewline
		\hline 
		$\{A,B\}$ & CVE-2022-46365 (Carol's collaboration was needed to fix any of the
		other two vulnerabilities) & 1\tabularnewline
		\hline 
		$\{A,C\}$ & CVE-2022-46365, CVE-2022-45801 & 2\tabularnewline
		\hline 
		$\{B,C\}$ & CVE-2022-45802 (other vulnerabilities would not be fixed without Alice) & 1\tabularnewline
		\hline 
		$\{A,B,C\}$ & all & 3\tabularnewline
		\hline 
	\end{tabular}
\end{table}

Computing the Shapley-value (with help of \cite{saavedra-nievesGameTheoryAllocationToolsCalculating2016}) gives the following fractions of how much Alice, Bob and Carol contributed to the avoidance of losses accordingly:
\begin{itemize}
	\item Alice: $\phi_A(v)=7/6$, normalized to $7/12$
	\item Bob: $\phi_B(v)= 1/6$, normalized to $1/12$
	\item Carol: $\phi_C(v)= 2/3$, normalized to $1/3$
\end{itemize}
The final payments due to contributions to security made by Alice, Bob and Carol, are then the normalized fractions of the Shapley-values, sharing the bonus $B$ to the three as follows:
\begin{itemize}
	\item Alice receives $\psi_A=\phi_A\cdot B\approx \$~12\,401.08$,
	\item Bob receives $\psi_B=\phi_B\cdot B\approx \$~1\,771.58$,
	\item Carol receives $\psi_C=\phi_C\cdot B\approx \$~7\,086.33$.
\end{itemize}

We conclude this example with the remark that all numbers given here are derived from actual data, yet a real life use of the model may require other sources of information. In light of this, the example is to be considered as artificial, and to showcase how the model could be used in a real life context. In an actual risk management process with developer incentivizations, design choices would have to be reconsidered in the given context.

\section{Discussion and Conclusion}\label{sec:conclusion}
This work builds upon the assumption that the host institution of a developer team does allocate a certain budget for security, but does not adopt assumptions on where this budget comes from. Retrieving this from increased pricing has, in prior work (see Section \ref{sec:related-work}) has found this to be a potentially undesirable strategy. We do not tackle the problem of how to obtain the budget, but emphasize that the bonus assign- and payment scheme described in Section \ref{sec:incentivization-mechanism} is agnostic of the size of the budget $B$. This means that the budget $B$ may come from any source, and is not necessarily linked to the particular product (or portfolio of products) that the developer team is concerned with.

Similarly, psychological factors are not covered yet, such as the possibility of a team member to actively argue to postpone security implementations in first place, only to later implement it alone to gain the full revenues. An a posteriori mitigation of such anti-coalitional behavior is by making the security revenues publicly announced among the team members, so that misbehavior would become exposed (publicly blaming such action). The effects of such a (psychological) precaution, as well as empirical studies on the intrinsic change of motivation to work for security in anticipation of bonuses paid for measurable efforts, merits its own research to be conducted and published as a follow-up work to this; the example from Section \ref{sec:worked-example} may serve as an initial blueprint to base this on.

From a practical perspective, the Shapley-value somewhat limits the scale of teams that we can reasonably analyze, which makes approximations \cite{clouseApproximatingPowerIndices2018} and alternative power indices \cite{Gimenez2017}  interesting to study. Especially the robustness against malicious team members that seek to secretly increase their revenues without increasing contributions is its own security problem inside the mechanism proposed here. Overall, the mechanism laid out in this work is conceptually simple to implement and offers explainability to (non-)receivers of bonuses, via the direct axiomatic grounding of the Shapley-value.

%
%
%



	\subsubsection{Acknowledgements} This work was supported by the Doctoral School ``Responsible Safe and Secure Robotic Systems Engineering (SEEROSE)'' at the Alpen-Adria-Universit\"at Klagenfurt. We also thank the anonymous reviewers for valuable suggestions that helped to improve the manuscript.
	
	%
	%
	%
	\bibliographystyle{splncs04}

\begin{thebibliography}{10}
\providecommand{\url}[1]{\texttt{#1}}
\providecommand{\urlprefix}{URL }
\providecommand{\doi}[1]{https://doi.org/#1}

\bibitem{adatEconomicIncentiveBased2018}
Adat, V., Dahiya, A., Gupta, B.B.: Economic incentive based solution against
  distributed denial of service attacks for {IoT} customers. In: 2018 {IEEE}
  {International} {Conference} on {Consumer} {Electronics} ({ICCE}). pp.~1--5.
  IEEE, Las Vegas, NV (Jan 2018). \doi{10.1109/ICCE.2018.8326280},
  \url{http://ieeexplore.ieee.org/document/8326280/}

\bibitem{apachefoundationApacheStreamParkIncubating2023}
{Apache Foundation}: Apache {StreamPark} {\textbar} {Incubating} (2023),
  \url{https://streampark.apache.org/}

\bibitem{augustDesigningUserIncentives2014}
August, T., August, R., Shin, H.: Designing user incentives for cybersecurity.
  Communications of the ACM  \textbf{57}(11),  43--46 (Oct 2014).
  \doi{10.1145/2629487}, \url{https://dl.acm.org/doi/10.1145/2629487}

\bibitem{augustNetworkSoftwareSecurity2006}
August, T., Tunca, T.I.: Network {Software} {Security} and {User} {Incentives}.
  Management Science  \textbf{52}(11),  1703--1720 (Nov 2006).
  \doi{10.1287/mnsc.1060.0568},
  \url{https://pubsonline.informs.org/doi/abs/10.1287/mnsc.1060.0568},
  publisher: INFORMS

\bibitem{cano-berlangaEnjoyingCooperativeGames2015}
Cano-Berlanga, S., Gimenez-Gomez, J.M., Vilella, C.: Enjoying cooperative
  games: {The} {R} package {GameTheory}. Working Paper No. 06; CREIP; Spain
  (Mar 2015)

\bibitem{ccconsortiumCommonCriteriaInformation2012a}
{CC Consortium}: Common {Criteria} for {Information} {Technology} {Security}
  {Evaluation} -- {Part} 3: {Security} assurance components (2012),
  \url{https://www.commoncriteriaportal.org/files/ccfiles/ccpart3v3.1r4.pdf}

\bibitem{ccconsortiumCommonCriteriaInformation2018}
{CC Consortium}: Common {Criteria} for {Information} {Technology} (2018),
  \url{https://www.commoncriteriaportal.org}

\bibitem{clouseApproximatingPowerIndices2018}
Clouse, D., Burke, D.: Approximating {Power} {Indices} to {Assess}
  {Cybersecurity} {Criticality}. In: Bushnell, L., Poovendran, R., Basar, T.
  (eds.) Decision and {Game} {Theory} for {Security}. pp. 346--365. Lecture
  {Notes} in {Computer} {Science}, Springer International Publishing, Cham
  (2018). \doi{10.1007/978-3-030-01554-1_20}

\bibitem{RePEc:aea:aecrev:v:85:y:1995:i:5:p:1187-1206}
Daughety, A.F., Reinganum, J.F.: Product safety: {Liability}, {R}\&{D}, and
  signaling. American Economic Review  \textbf{85}(5),  1187--1206 (Dec 1995),
  \url{https://ideas.repec.org/a/aea/aecrev/v85y1995i5p1187-1206.html}

\bibitem{duenasPercevalSoftwareProject2018}
Due\~{n}as, S., Cosentino, V., Robles, G., Gonzalez-Barahona, J.M.: Perceval:
  {Software} {Project} {Data} at {Your} {Will}. In: 2018 {IEEE}/{ACM} 40th
  {International} {Conference} on {Software} {Engineering}: {Companion}
  ({ICSE}-{Companion}). pp.~1--4 (May 2018), iSSN: 2574-1934

\bibitem{eetenEconomicsMalwareSecurity2008}
Eeten, M.J.G.v., Bauer, J.M.: Economics of {Malware}: {Security} {Decisions},
  {Incentives} and {Externalities}. Tech. rep., OECD, Paris (May 2008).
  \doi{10.1787/241440230621},
  \url{https://www.oecd-ilibrary.org/content/paper/241440230621}

\bibitem{european_commission_regulation_2016}
{European Commission}: Regulation ({EU}) 2016/679 of the {European}
  {Parliament} and of the {Council} of 27 {April} 2016 on the protection of
  natural persons with regard to the processing of personal data and on the
  free movement of such data, and repealing {Directive} 95/46/{EC} ({General}
  {Data} {Protection} {Regulation}) ({Text} with {EEA} relevance) (2016),
  \url{https://eur-lex.europa.eu/eli/reg/2016/679/oj}, tex.added-at:
  2020-08-20T11:33:21.000+0200 tex.biburl:
  https://www.bibsonomy.org/bibtex/243a2175512dc8b9d8855fa7a763cdc3e/zotero
  tex.interhash: c7b667cac6031282160a9e94d5a118f8 tex.intrahash:
  43a2175512dc8b9d8855fa7a763cdc3e tex.timestamp: 2020-08-20T11:33:21.000+0200

\bibitem{firstforumofincidentresponseandsecurityteamsCVSSV3Specification}
{FIRST (Forum of Incident Response and Security Teams)}: {CVSS} v3.1
  {Specification} {Document},
  \url{https://www.first.org/cvss/specification-document}

\bibitem{garayRationalProtocolDesign2013}
Garay, J., Katz, J., Maurer, U., Tackmann, B., Zikas, V.: Rational {Protocol}
  {Design}: {Cryptography} against {Incentive}-{Driven} {Adversaries}. In: 2013
  {IEEE} 54th {Annual} {Symposium} on {Foundations} of {Computer} {Science}.
  pp. 648--657. IEEE (2013). \doi{10.1109/FOCS.2013.75}, event-place:
  Piscataway, NJ

\bibitem{Gimenez2017}
Gim\'enez, J., Puente, A.: A new procedure to calculate the owen value. In:
  Proceedings of the 6th {International} {Conference} on {Operations}
  {Research} and {Enterprise} {Systems} ({ICORES} 2017). pp. 228--233 (Jan
  2017). \doi{10.5220/0006113702280233}

\bibitem{gordonRationalSecretSharing2006}
Gordon, S.D., Katz, J.: Rational {Secret} {Sharing}, {Revisited}. In: Prisco,
  R., Yung, M. (eds.) Security and {Cryptography} for {Networks}.5th
  {International} {Conference}, {SCN} 2006, {Maiori}, {Italy}, {September} 6-8,
  2006, {Proceedings}, Lecture {Notes} in {Computer} {Science}, vol.~4116, pp.
  229--241. Springer-Verlag GmbH, Berlin Heidelberg (2006).
  \doi{\url{10.1007/11832072_16}},
  \url{\url{https://doi.org/10.1007/11832072_16}}

\bibitem{gousiosGhtorrentGhtorrentOrg2020}
Gousios, G.: ghtorrent/ghtorrent.org (2020),
  \url{https://github.com/ghtorrent/ghtorrent.org/blob/51890965af72da85bdd0954a50ef1a71603fb4e7/faq.md},
  original-date: 2013-02-23T16:03:24Z

\bibitem{haldermanStrengthenSecurityChange2010}
Halderman, J.A.: To {Strengthen} {Security}, {Change} {Developers}'
  {Incentives}. IEEE Security \& Privacy  \textbf{8}(2),  79--82 (Mar 2010).
  \doi{10.1109/MSP.2010.85}, conference Name: IEEE Security \& Privacy

\bibitem{howardWritingSecureCode2003}
Howard, M., LeBlanc, D.: Writing {Secure} {Code}. Microsoft Press, 2nd edn.
  (2003)

\bibitem{huangRepChainReputationBasedSecure2021}
Huang, C., Wang, Z., Chen, H., Hu, Q., Zhang, Q., Wang, W., Guan, X.:
  {RepChain}: {A} {Reputation}-{Based} {Secure}, {Fast}, and {High} {Incentive}
  {Blockchain} {System} via {Sharding}. IEEE Internet of Things Journal
  \textbf{8}(6),  4291--4304 (Mar 2021). \doi{10.1109/JIOT.2020.3028449},
  conference Name: IEEE Internet of Things Journal

\bibitem{informationtechnologylaboratoryNVDHome2023}
{Information Technology Laboratory}: {NVD} - {Home} (2023),
  \url{https://nvd.nist.gov/}

\bibitem{informationtechnologylaboratoryNVDHome}
{Information Technology Laboratory}: {NVD} - {Home} (2023),
  \url{https://nvd.nist.gov/}, [retrieved: May 12, 2023]

\bibitem{organisationISOIEC270002016}
{International Organization for Standardization}: {ISO}/{IEC} 27001 --
  {Information} technology -- {Security} techniques -- {Information} security
  management systems -- {Requirements} (2013),
  \url{http://www.iso.org/iso/iso27001}

\bibitem{isoISOIEC270022023}
{ISO}: {ISO}/{IEC} 27002 controls catalogue (2023),
  \url{https://www.iso27001security.com/html/27002.html}

\bibitem{kawachiGeneralConstructionsRational2017}
Kawachi, A., Okamoto, Y., Tanaka, K., Yasunaga, K.: General {Constructions} of
  {Rational} {Secret} {Sharing} with {Expected} {Constant}-{Round}
  {Reconstruction}. The Computer Journal  \textbf{60}(5),  711--728 (Apr 2017).
  \doi{10.1093/comjnl/bxw094}, \url{https://doi.org/10.1093/comjnl/bxw094}

\bibitem{liCreditCoinPrivacyPreservingBlockchainBased2018}
Li, L., Liu, J., Cheng, L., Qiu, S., Wang, W., Zhang, X., Zhang, Z.:
  {CreditCoin}: {A} {Privacy}-{Preserving} {Blockchain}-{Based} {Incentive}
  {Announcement} {Network} for {Communications} of {Smart} {Vehicles}. IEEE
  Transactions on Intelligent Transportation Systems  \textbf{19}(7),
  2204--2220 (Jul 2018). \doi{10.1109/TITS.2017.2777990},
  \url{https://ieeexplore.ieee.org/document/8267113/}

\bibitem{liuIncentivebasedModelingInference2005a}
Liu, P., Zang, W., Yu, M.: Incentive-based modeling and inference of attacker
  intent, objectives, and strategies. ACM Transactions on Information and
  System Security  \textbf{8}(1),  78--118 (Feb 2005).
  \doi{10.1145/1053283.1053288},
  \url{https://dl.acm.org/doi/10.1145/1053283.1053288}

\bibitem{lvResearchProfitDistribution2013}
Lv, X., Zhao, S.: Research on {Profit} {Distribution} of {Software}
  {Outsourcing} {Alliances} {Based} on the {Improved} {Shapley} {Value}
  {Model}. Cybernetics and Information Technologies
  \textbf{13}(Special-Issue),  100--109 (Dec 2013).
  \doi{10.2478/cait-2013-0041},
  \url{https://sciendo.com/article/10.2478/cait-2013-0041}

\bibitem{sets2}
Meyer, D., Hornik, K.: Generalized and customizable sets in {R}. Journal of
  Statistical Software  \textbf{31}(2),  1--27 (2009).
  \doi{10.18637/jss.v031.i02}

\bibitem{sets1}
Meyer, D., Hornik, K.: sets: Sets, Generalized Sets, Customizable Sets and
  Intervals (2023), \url{https://CRAN.R-project.org/package=sets}, r package
  version 1.0-24

\bibitem{munchWegeZurRisikobewertung2012}
M\"unch, I.: Wege zur {Risikobewertung}. pp. 326--337. syssec (2012)

\bibitem{saavedra-nievesGameTheoryAllocationToolsCalculating2016}
Saavedra-Nieves, A.: {GameTheoryAllocation}: {Tools} for calculating
  allocations in game theory. manual (2016),
  \url{https://CRAN.R-project.org/package=GameTheoryAllocation}

\bibitem{starmerDevelopmentsNonExpectedUtility2000}
Starmer, C.: Developments in {Non}-{Expected} {Utility} {Theory}: {The} {Hunt}
  for a {Descriptive} {Theory} of {Choice} under {Risk}. Journal of Economic
  Literature  \textbf{38}(2),  332--382 (2000),
  \url{http://www.jstor.org/stable/2565292}

\bibitem{tverskyFramingDecisionsPsychology1985}
Tversky, A., Kahneman, D.: The {Framing} of {Decisions} and the {Psychology} of
  {Choice}. In: Covello, V.T., Mumpower, J.L., Stallen, P.J.M., Uppuluri,
  V.R.R. (eds.) Environmental {Impact} {Assessment}, {Technology} {Assessment},
  and {Risk} {Analysis}, pp. 107--129. {NATO} {ASI} {Series}, {Series} {G},
  Springer, Berlin and Heidelberg (1985). \doi{10.1007/978-3-642-70634-9_6}

\bibitem{tverskyRationalChoiceFraming1989}
Tversky, A., Kahneman, D.: Rational {Choice} and the {Framing} of {Decisions}.
  In: Karpak, B., Zionts, S. (eds.) Multiple {Criteria} {Decision} {Making} and
  {Risk} {Analysis} {Using} {Microcomputers}, pp. 81--126. {NATO} {ASI}
  {Series}, {Series} {F}, Springer, Berlin and Heidelberg (1989).
  \doi{10.1007/978-3-642-74919-3_4}

\bibitem{widup2015VerizonData2015}
Widup, S., Rudis, B., Hylender, D., Spitler, M., Thompson, K., Baker, W.,
  Bassett, G., Karambelkar, B., Brannon, S., Kennedy, D., Jacobs, J.: 2015
  {Verizon} {Data} {Breach} {Investigations} {Report}. Tech. rep., Verizon (Apr
  2015). \doi{10.13140/RG.2.1.4205.5768},
  \url{https://www.researchgate.net/publication/289254638_2015_Verizon_Data_Breach_Investigations_Report}

\bibitem{xieImprovedShapleyValue2020}
Xie, W., Yu, X., Zhang, Y., Wang, H.: An {Improved} {Shapley} {Value} {Benefit}
  {Distribution} {Mechanism} in {Cooperative} {Game} of {Cyber} {Threat}
  {Intelligence} {Sharing}. In: {IEEE} {INFOCOM} 2020 - {IEEE} {Conference} on
  {Computer} {Communications} {Workshops} ({INFOCOM} {WKSHPS}). pp. 810--815.
  IEEE, Toronto, ON, Canada (Jul 2020).
  \doi{10.1109/INFOCOMWKSHPS50562.2020.9162739},
  \url{https://ieeexplore.ieee.org/document/9162739/}

\end{thebibliography}

\end{document}